\documentclass[aps,pr,reprint,superscriptaddress]{revtex4-2}

\usepackage[separate-uncertainty=true]{siunitx}

\usepackage{graphicx} 
\usepackage{amsmath}

\graphicspath{{figures/}}
\renewcommand{\vec}[1]{\mathbf{#1}}
\newcommand*{\abs}[1]{\left \vert #1 \right \vert}

\begin{document}

\title{Holographic characterization and tracking
of colloidal dimers in the effective-sphere
approximation} 

\author{Lauren E. Altman}
\affiliation{Department of Physics and Center for Soft Matter
  Research, New York University, New York, NY 10003, USA}

\author{Rushna Quddus}
\affiliation{Department of Chemistry,
  New York University, New York, NY 10003, USA}

\author{Fook Chiong Cheong}
\affiliation{Spheryx, Inc., 330 E.\ 38th Street, \#48J, New York, NY
  10016, USA}

\author{David G. Grier}
\affiliation{Department of Physics and Center for Soft Matter
  Research, New York University, New York, NY 10003, USA}

\begin{abstract}
An in-line hologram of a colloidal sphere can be
analyzed with the Lorenz-Mie theory of light scattering
to measure the sphere's three-dimensional position
with nanometer-scale precision while also
measuring its diameter and refractive index with
part-per-thousand precision.
Applying the same technique to aspherical or inhomogeneous particles yields the position,
diameter and
refractive index of an effective sphere that represents an average
over the particle's geometry and composition.
This effective-sphere
interpretation has been applied successfully to porous, dimpled and
coated spheres, as well as to fractal clusters of nanoparticles, all of whose inhomogeneities
appear on length scales smaller than the wavelength of light.
Here, we combine numerical and
experimental studies to investigate effective-sphere characterization
of symmetric dimers of micrometer-scale spheres, a class of aspherical objects that appear commonly in real-world dispersions.
Our studies demonstrate that the effective-sphere interpretation
usefully identifies dimers in
holographic characterization studies of monodisperse
colloidal spheres.
The effective-sphere estimate for a dimer's
axial position closely
follows the ground truth for its center of mass.
Trends in the effective-sphere 
diameter and refractive index,
furthermore, can be used to measure
a dimer's three-dimensional orientation.
When applied to colloidal dimers transported in a
Poiseuille flow, the estimated orientation distribution is consistent with expectations for
Brownian particles undergoing Jeffery orbits.
\end{abstract}

\maketitle

\section{Introduction}

\begin{figure*}
    \centering
    \includegraphics[width=0.9\textwidth]{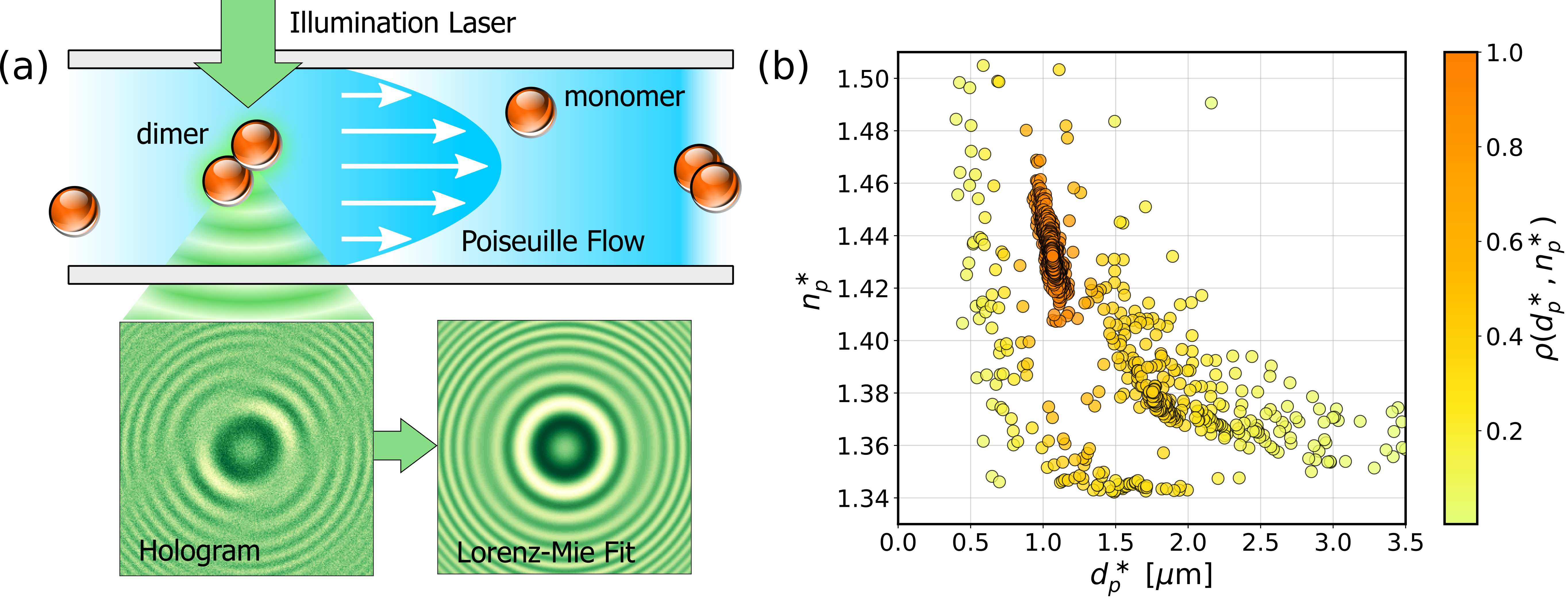}
    \caption{(a) Schematic representation of holographic characterization
    of colloidal spheres and clusters in a microfluidic channel.
    Particles, including clusters, passing through the 
    illuminating laser beam scatter light that interferes with
    the rest of the laser beam to create a hologram. Fitting a hologram to
    a generative model based on the
    Lorenz-Mie theory of light scattering yields measurements
    of the particle's effective diameter,
    $d_p^\ast$, and refractive index, $n_p^\ast$.
    (b) Scatter plot of single-particle
    properties measured for a
    nominally monodisperse 
    sample of micrometer-diameter silica spheres that have been
    functionalized for a medical diagnostic test.
    Each plotted point represents the measured 
    properties of
    one particle and is colored by the relative
    probability density of observations,
    $\rho(d_p^\ast, n_p^\ast)$. Clusters of points
    are likely to represent distinct types
    of particles, with at least three clusters
    being evident here.}
    \label{fig:schematic}
\end{figure*}

Holographic particle characterization
\cite{lee_characterizing_2007} is a fast and robust 
technique for measuring the size and
refractive index of colloidal spheres and has
a wide variety of applications in basic research
\cite{xiao_multidimensional_2010,wang_holographic_2016,middleton_optimizing_2019,zagzag_holographic_2020},
industrial materials analysis \cite{wang_holographic_2016-1}
and medical diagnostics \cite{zagzag_holographic_2020,snyder_holographic_2020}.
As illustrated in Fig.~\ref{fig:schematic}(a), holographic particle
characterization uses in-line 
holographic video microscopy \cite{sheng_digital_2006,lee_characterizing_2007}
to record holograms of individual colloidal particles
and then fits those holograms to predictions of the
Lorenz-Mie theory of light scattering
\cite{bohren_absorption_1983,mishchenko_scattering_2002,gouesbet_generalized_2011}
to track the the particle in three dimensions and
to measure its size and refractive index.
Originally developed for analyzing homogeneous isotropic spheres,
holographic particle characterization also has been applied
to dimpled spheres \cite{hannel_holographic_2015}, 
porous spheres \cite{cheong_holographic_2011,odete_role_2020},
coated spheres \cite{altman_interpreting_2020},
and fractal aggregates of colloidal nanoparticles
\cite{wang_holographic_2016,cheong_holographic_2017,fung_computational_2019}.
In all cases, the particles are analyzed with 
a generative model based on the theory of
light scattering by homogeneous spheres.
The properties obtained in this way
therefore should be interpreted as parameters
of an effective-sphere model.
Effective-sphere properties then can be related to actual properties
such as porosity \cite{cheong_holographic_2011,odete_role_2020}
or fractal dimension \cite{wang_holographic_2016,fung_computational_2019}
through effective-medium theory \cite{cheong_holographic_2011,markel_introduction_2016,odete_role_2020}.

The success of the effective-sphere model hinges on the
assumption that inhomogeneities in the particles' structure
or composition are evenly distributed 
throughout the volume of the
particle on length scales smaller than
the wavelength of light.
This is not the case, for example, when 
micrometer-scale colloidal spheres aggregate into
dimers, trimers and similar small clusters.
Holograms formed by such clusters can be analyzed
quantitatively using suitable generative models
for light scattering by clusters \cite{fung_imaging_2012,fung_holographic_2013}.
Such detailed analysis, however, is orders of magnitude
slower than the corresponding effective-sphere measurement.
Small clusters, moreover, figure prominently
in holographic characterization measurements of
real-world colloidal dispersions, such as the example
in Fig.~\ref{fig:schematic}(b).
The ability to identify and interpret holograms 
of such clusters within the effective-sphere
framework therefore would create
immediate applications in areas as diverse as
semiconductor processing 
\cite{cheong_holographic_2017}
and medical testing \cite{snyder_holographic_2020}.

Here, we use numerical studies to assess
the utility
of the effective-sphere model for interpreting
holograms of small colloidal
clusters composed of micrometer-scale spheres.
These studies demonstrate that small clusters 
can be clearly identified in dispersions of
monodisperse colloidal spheres.
The ability to identify and count colloidal
dimers 
provides a basis for measuring their relative
abundance,
which is useful in such applications as detecting
agglutination in bead-based medical assays
\cite{zagzag_holographic_2020,snyder_holographic_2020}.

When effective-sphere analysis is applied to dimers,
the observed distribution of single-particle properties
is found to be parameterized by the dimers' 
orientation relative to the optical axis.
Holographic characterization in the effective-sphere
approximation therefore enables real-time
tracking of colloidal dimers' orientations.
We apply this technique to experimental studies
of silica dimers transported in a Poiseuille flow 
down a microfluidic channel and demonstrate that
the observed distribution of dimer orientations is
consistent with expectations for Jeffery orbits.
For symmetric colloidal dimers, therefore,
the effective-sphere model appears to offer
quantitative insights into the particles'
trajectories and orientations.

\section{Holographic particle characterization}
\label{sec:holographicparticlecharacterization}

Holographic particle characterization is 
an application of
in-line holographic video microscopy \cite{sheng_digital_2006},
in which the sample is illuminated with a
collimated laser beam.
A colloidal particle entering the beam 
scatters some of the light,
and the scattered light interferes with the rest of the beam in the focal plane
of a conventional optical microscope.
The microscope magnifies the interference
pattern and
relays it to a video camera that records
its intensity.
Each image in the resulting
video stream
is a hologram of the particles
in the observation volume
and encodes information
about their three-dimensional
positions, as well as their sizes, shapes
and compositions.
This information can be
extracted by fitting recorded holograms
to a generative model
for the image-formation process.
The standard implementation \cite{lee_characterizing_2007}
models the incident beam as a unit-amplitude
monochromatic plane wave at frequency $\omega$,
whose electric field,
\begin{equation}
\label{eq:incidentbeam}
    \vec{E}_0(\vec{r}, t)
    =
    e^{i k z} e^{-i \omega t} \, \hat{x},
\end{equation}
is linearly polarized along $\hat{x}$
and propagates along $\hat{z}$.
The wavenumber, $k = 2 \pi n_m/\lambda$,
depends on the laser's vacuum wavelength, $\lambda$,
and the refractive index of the medium,
$n_m$, and is related to the
frequency through the standard dispersion
relation, $\omega = c k$, where $c$ is
the speed of light in the medium.

This plane wave illuminates a particle
located at $\vec{r}_p$ relative to the
center of the microscope's focal plane,
which scatters some of the light.
The time-averaged
intensity pattern recorded by the camera
may be modeled as a superposition
of the incident and scattered waves,
\begin{equation}
\label{eq:b}
    b(\vec{r})
    =
    \abs{
    \hat{x} + 
    e^{-i k z_p} \,
    \vec{f}_s(k(\vec{r} - \vec{r}_p))
    }^2,
\end{equation}
where $\vec{f}_s(k\vec{r})$ is the Lorenz-Mie
scattering function for the particle \cite{bohren_absorption_1983,mishchenko_scattering_2002,gouesbet_generalized_2011}.
Experimentally recorded holograms
are corrected for the dark count of the camera
and normalized by the intensity distribution
of the illumination
to facilitate comparison with Eq.~\eqref{eq:b}.

For the particular case of a spherical scatterer,
the Lorenz-Mie scattering function
is parametrized by
the sphere's diameter, $d_p$, and refractive
index, $n_p$.
Analyzing a single-particle hologram with
Eq.~\eqref{eq:b} involves optimizing these two parameters
in addition to the particle's three-dimensional position,
$\vec{r}_p$.
The only calibration constants required for this
analysis are the vacuum wavelength
of the laser, $\lambda$, the magnification
of the microscope and the refractive index of
the medium, $n_m$, all of which can be measured
independently.
For the studies that follow,
we choose $\lambda = \SI{445}{\nm}$
and a system magnification of
\SI{120}{\nm\per pixel} for compatibility
with commercial
holographic particle characterization
instruments
(Spheryx, Inc., xSight).
We furthermore assume that the medium has the
refractive index of water at the imaging
wavelength, $n_m = \num{1.340}$.

Standard software implementations
of holographic particle characterization
\cite{wiscombe_improved_1980,yang_improved_2003,pena_scattering_2009,krishnatreya_fast_2014,altman_catch_2020}
can localize and characterize a sphere
within 50 milliseconds
on a desktop computer workstation.
Validation experiments on colloidal size standards
demonstrate that each such fit
can resolve the diameter of a micrometer-scale
sphere with a precision of \SI{\pm 2}{\nm} \cite{lee_holographic_2007}. 
Measurements on emulsion droplets demonstrate
precision and reproducibility in the refractive
index of \num{\pm 1} part per thousand \cite{shpaisman_holographic_2012}.
The former is fine enough to
detect the formation of a molecular coating on a bead
through the associated change in the bead's diameter
\cite{cheong_flow_2009,zagzag_holographic_2020}.
The latter is useful for distinguishing different
types of beads on the basis of their
composition \cite{yevick_machine-learning_2014}.

\section{The Effective-Sphere Model}
\label{sec:effectivespheremodel}

\subsection{Inhomogeneous Spheres}

The standard Lorenz-Mie analysis treats the
scatterer as an isotropic homogeneous sphere.
A porous sphere, by contrast, is composed of
a matrix of refractive index $n_0$ arranged
at volume fraction $\Phi$ within the body of the
particle. The remainder of the sphere's
volume is
filled with the medium at refractive index $n_m$.
This inhomogeneous particle still may be treated
as a homogeneous medium if its inhomogeneities
are distributed uniformly through its volume 
on a length scale smaller than the wavelength 
of light.
In that case, its effective refractive index,
$n_p^\ast$, may be related to $n_0$, $n_m$ and $\Phi$
through Maxwell Garnett effective medium theory
\cite{markel_introduction_2016,cheong_holographic_2011,odete_role_2020}:
\begin{subequations}
\label{eq:effectivesphere}
\begin{equation}
    n_p^\ast
    =
    n_m
    \sqrt{\frac{1 + 2 \Phi L(n_0/n_m)}{1 - \Phi L(n_0/n_m)}},
\end{equation}
where the Lorentz-Lorenz factor is
\begin{equation}
    L(m) = \frac{m^2 - 1}{m^2 + 2}.
\end{equation}
\end{subequations}

The effective-sphere model described by
Eq.~\eqref{eq:effectivesphere}
has been found to 
agree quantitatively with experimental results for
mesoporous colloidal spheres
\cite{odete_role_2020},
which should satisfy all of the assumptions underlying the
model.
More surprisingly, Eq.~\eqref{eq:effectivesphere}
also yields quantitative results for fractal
clusters of dielectric nanoparticles
\cite{wang_holographic_2016,fung_computational_2019}
even though those clusters are neither uniformly
porous nor ideally spherical. 
These experimental \cite{wang_holographic_2016}
and simulation studies \cite{fung_computational_2019}
suggest that the effective diameter,
$d_p^\ast$, that emerges
from Lorenz-Mie analysis roughly bounds
the irregular cluster and 
that the effective refractive index,
$n_p^\ast$, is consistent with the volume fraction of
material, $\Phi$, within that bounding sphere.

\subsection{Symmetric Colloidal Dimers}

The hologram of a colloidal cluster
could be analyzed by suitably generalizing
the scattering function, $\vec{f}_s(k\vec{r})$,
to account for the cluster's 
composition, geometry and orientation
\cite{fung_imaging_2012}.
Introducing these additional adjustable
parameters, however, reduces the likelihood
that the fits will converge successfully,
increases the measurement's sensitivity to
imaging imperfections and very substantially
increases the measurement time \cite{barkley_holographic_2019}.
In many cases of interest, furthermore, the
detailed geometry of a small colloidal cluster
is of less interest than the ability to
differentiate clusters from monomeric spheres.
This is the case, for example, in medical diagnostic
tests where agglutination of functionalized beads can
indicate successful detection of the
target analyte \cite{alves_rapid_2020}.
We therefore seek to understand how the standard
effective-sphere analysis 
accommodates the geometry and
orientation of small colloidal clusters.

We approach this problem through numerical
experiments, generating
synthetic holograms
of colloidal clusters then applying 
effective-sphere
analysis to obtain the clusters'
effective positions, diameters and refractive indexes.
These effective parameters then can be
interpreted in light of the ground-truth
properties of the clusters, thereby providing
guidance for interpreting experimental
particle-characterization data.

We compute the scattering functions
for colloidal clusters
with cluster T matrix theory 
\cite{mackowski_96,mackowski_2011}
using the \texttt{HoloPy} API 
\cite{barkley_holographic_2019},
then apply Eq.~\eqref{eq:b}
to compute ideal holograms,
and finally add 
\SI{5}{\percent} Gaussian noise
to simulate
experimental recordings.
We analyze those synthetic holograms with
the \texttt{pylorenzmie} implementation
of holographic particle characterization.
Results obtained by analyzing synthetic holograms
then can be compared directly
with experimental observations on
colloidal dispersions that are known
to include aggregates.

\begin{figure}
    \centering
    \includegraphics[width=0.5\columnwidth]{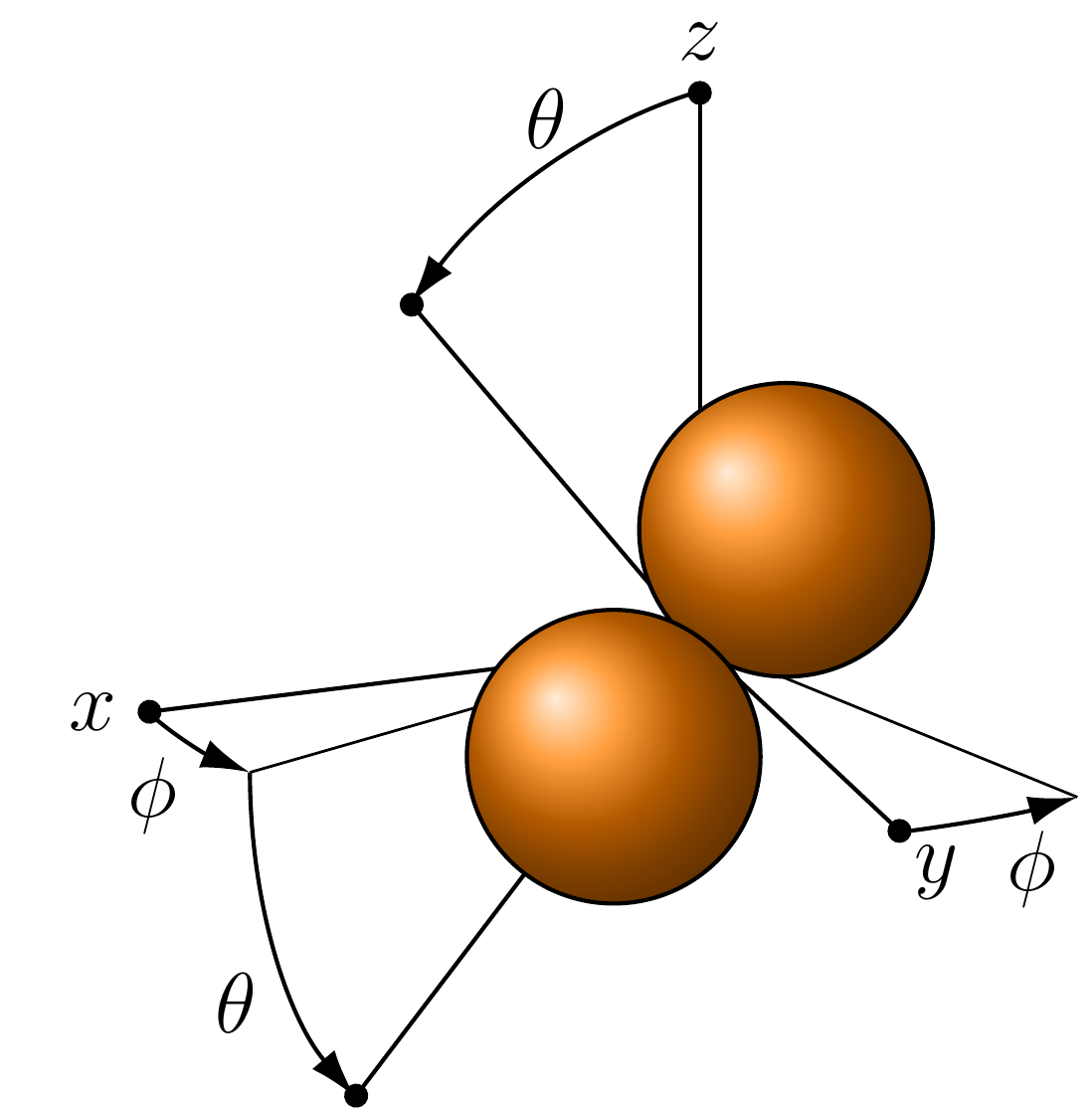}
    \caption{Geometry of symmetric dimers used
    for computing cluster holograms.
    A pair of spheres initially is aligned with the
    $\hat{x}$ axis and then is rotated through two Euler angles, $\phi$ and $\theta$.}
    \label{fig:geometry}
\end{figure}

\begin{figure*}
    \centering
    \includegraphics[width=0.8\textwidth]{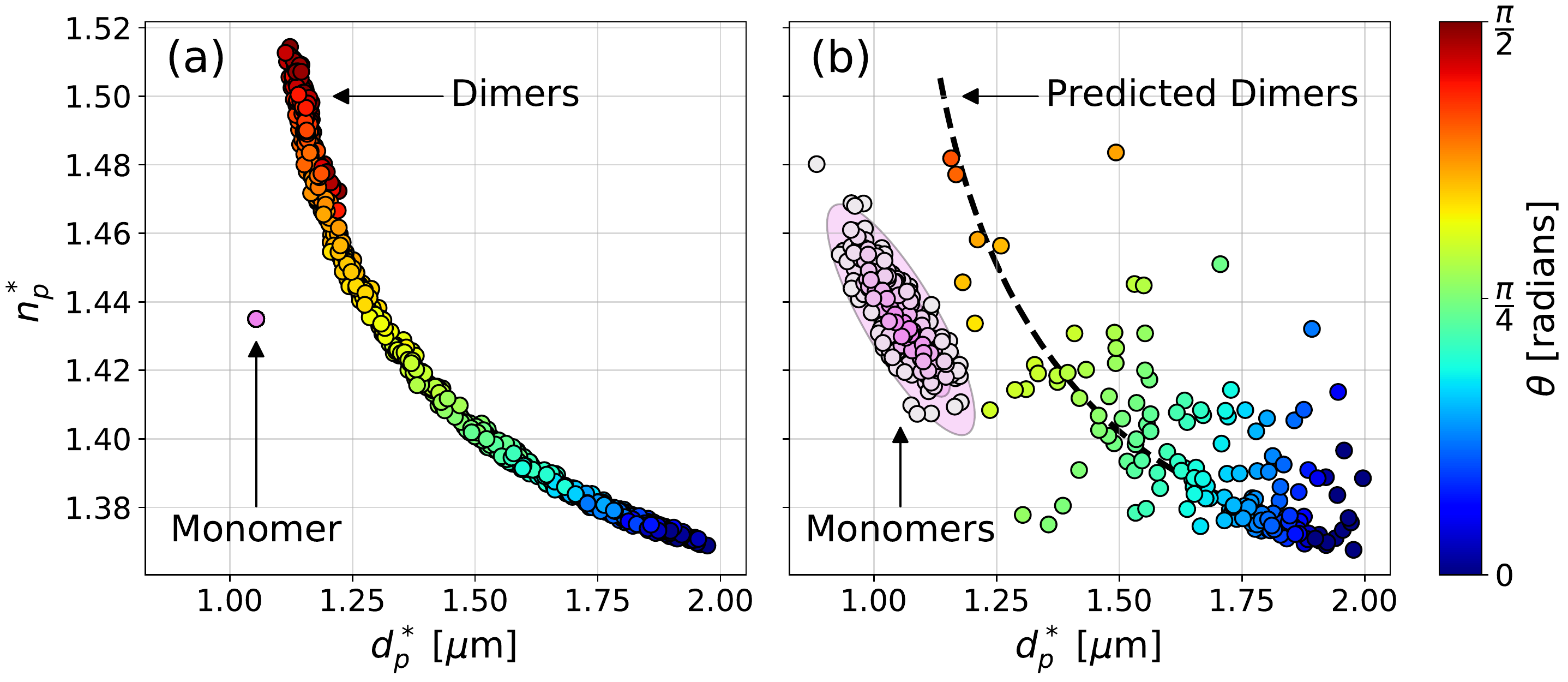}
    \caption{Effective-sphere interpretation 
    of a (a) simulated and (b) experimental holograms
    of colloidal monomers and dimers.
    Simulations are performed for symmetric dimers
    composed of  silica spheres 
    ($n_0 = \num{1.435}$) with diameter
    $d_0 = \SI{1.054}{\um}$ at varying polar angle $\theta$. Individual points are colored
    by $\theta$. The values for $d_0$ and $n_0$
    are selected to mimic the experimental system.
    The dimers' effective diameters, $d_p^*(\theta)$,
    and refractive indexes, $n_p^*(\theta)$, are fitted to the 
    phenomenological model from Eq.~\eqref{eq:dimerfits}
    that is plotted as the dashed black curve
    in (b). Holographic characterization data
    are segmented into monomers and dimers.
    Data points identified with dimers are colored by
    the angle $\theta$ of closest approach to
    the predicted dimer curve.}
    \label{fig:dimers_theta_fits}
\end{figure*}

Figure~\ref{fig:geometry} shows the geometry used
for computing synthetic holograms of colloidal 
dimers.
A symmetric dimer centered at the origin can be 
parameterized by the monomers' diameter, $d_p$, 
and the Euler angles, $\phi$ and $\theta$.
The polar angle $\theta = 0$ corresponds to the dimer 
lying with its axis in the imaging plane.
At $\theta = \pi/2$, the dimer is aligned
with the optical axis.
The azimuthal angle $\phi = 0$ aligns the
dimer with the light's axis of polarization.
With these definitions, 
the centers of the two monomers are situated at 
$\vec{r}_\pm = \pm \frac{1}{2} d_p \, \hat{n}$
relative to the dimer's center, $\vec{r}_p$, where 
$\hat{n} = 
(\cos\phi \cos\theta, \sin\phi \cos\theta, \sin\theta)$ 
is the dimer's orientation unit vector.
The dimer's description is completed with
choices for the spheres' refractive index,
$n_p$, and the axial position, $z_p$, of
the dimer's center relative to the microscope's
focal plane.
A synthetic hologram defined by the
set of parameters,
$(d_p, n_p, z_p, \theta, \phi)$,
is analyzed with Eq.~\eqref{eq:b}
to obtain the effective-sphere descriptors,
$d_p^\ast$ and $n_p^\ast$, as well as an
estimate for the dimer's axial position,
$z_p$.

\subsection{Effective Dimer Size and Refractive Index}

Results of a typical numerical study are presented in
Fig.~\ref{fig:dimers_theta_fits}(a).
Each point in this scatter plot
represents the effective diameter and 
refractive index of a single particle,
either a monomeric sphere
or a randomly oriented dimer.
The spheres' properties
are selected to mimic 
the monodisperse silica probe beads used
for holographic immunoassays \cite{snyder_holographic_2020}.
Values obtained for the spheres' properties
are found to be independent of axial position over the 
range 
$\SI{20}{\um} < z_p < \SI{120}{\um}$, with the lower
bound being set by the
spatial resolution of
the imaging grid
and the upper bound
by the images' 
signal-to-noise ratio.
The extended cluster of points 
in Fig.~\ref{fig:dimers_theta_fits}(a)
represents results for
colloidal dimers with random orientations in the
range $0 \le \theta \le \frac{\pi}{2}$
and $0 \le \phi \le \pi$,
and with axial positions in the range $\SI{50}{\mu m} \le z_p \le \SI{95}{\mu m}$.
Each point in the dimer distribution is colored
by the ground truth value for $\theta$.

As might be expected, 
dimers lying in the imaging plane appear
nearly twice as large as their component spheres.
This is consistent with previous applications of
effective-sphere analysis to irregular colloidal
particles \cite{wang_holographic_2016,fung_computational_2019}
that found the estimated parameters to correspond to an
average over a bounding sphere.
The effective refractive index of the in-plane dimer therefore is
closer to that of the medium because the volume
fraction of the silica spheres within the effective sphere
is just $\Phi = \frac{1}{4}$.
Equation~\eqref{eq:effectivesphere} then sets the
expected lower bound for the observed refractive index
to be $n_p^\ast > \num{1.36}$.

The aspherical structure of a dimer is far less
apparent when the dimer is aligned with the optical
axis, $\theta = \frac{\pi}{2}$.
An aligned dimer scatters more light than a single
sphere and thus appears to have a higher refractive index.
Its effective diameter, however, is only slightly
larger than that of a monomer.
The effective-sphere model from Sec.~\ref{sec:effectivespheremodel}
does not account for these observations presumably because
the dimer's inhomogeneities appear on a scale larger
than the wavelength of light, which means that the particle's
orientation strongly influences its light-scattering properties.
The apparent agreement for the in-plane dimer
is all the more remarkable when viewed in this light.

It should be noted at this point that the identity of
any individual colloidal particle as a dimer would 
not be evident on the basis of these single-particle characterization
measurements. The distribution of particle characterization
data in the $(d_p^\ast, n_p^\ast)$ plane, however, has a 
characteristic shape that can be recognized in the broader
distribution of a mixed sample,
such as the example in Fig.~\ref{fig:schematic}(b).
If we identify the principal peak in those data with
a population of monodisperse spheres, then the set of
larger particles at lower refractive index trace out
a curve that can be recognized as arising from dimers.

\subsection{Dependence on Dimer Orientation}

\begin{figure}
    \centering
    \includegraphics[width=\columnwidth]{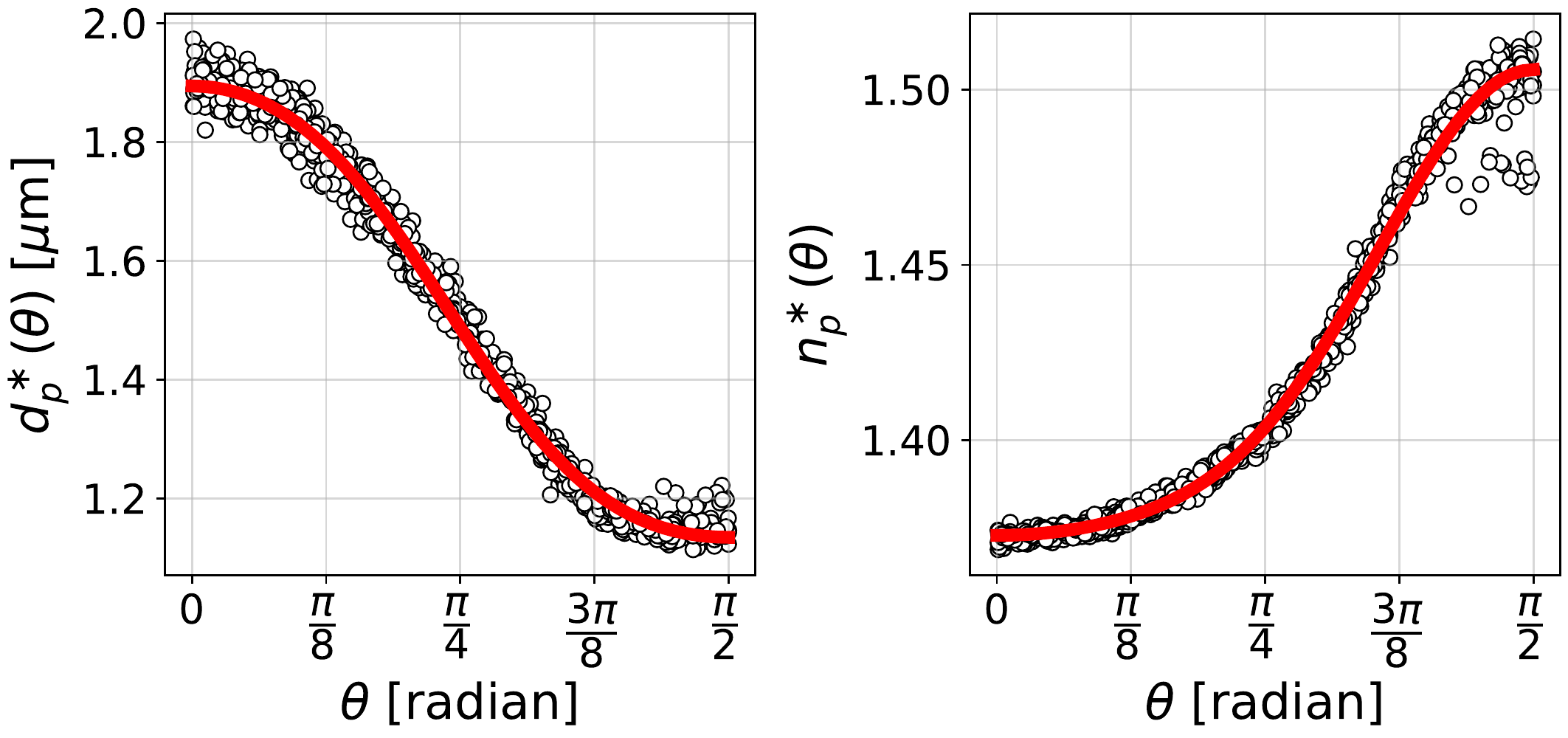}
    \caption{Dependence
    of the effective-sphere
    diameter, $d_p^\ast(\theta)$, and refractive index,
    $n_p^\ast(\theta)$, on dimer orientation, $\theta$. Discrete points are the data for simulated
    dimers
    plotted in Fig.~\ref{fig:dimers_theta_fits}(a).
    Solid (red) curves are fits to Eq.~\eqref{eq:dimerfits}.}
    \label{fig:dimerfits}
\end{figure}

The results in Fig.~\ref{fig:dimers_theta_fits}(a)
suggest that the distribution of effective properties
recorded for dimers is simply parameterized by $\theta$
and does not depend significantly on $\phi$.
Lacking a theoretical model for $d_p^\ast(\theta)$ and
$n_p^\ast(\theta)$, we instead fit the
measured effective-sphere results to
polynomials in $\sin^2(\theta)$,
\begin{subequations}
\label{eq:dimerfits}
\begin{align}
    d_p^\ast(\theta) & = \sum_{j = 0}^N d_j^\ast \, \sin^{2j}(\theta) \\
    n_p^\ast(\theta) & = \sum_{j = 0}^N n_j^\ast \, \sin^{2j}(\theta),
\end{align}
\end{subequations}
which respect the symmetries of the system and should
yield the effective-sphere predictions for
the in-plane properties,
$d_0^\ast$ and $n_0^\ast$.

The results presented in Fig~\ref{fig:dimerfits} are obtained
with $N = 3$ and yield
$d_0^\ast = \SI{1.9}{\um}$ and $n_0^\ast = \num{1.37}$.
The fit value for the aligned dimers' diameter 
is slightly smaller than the diameter of the bounding
sphere, and the fit value for the refractive index
is correspondingly higher.
The computed data in Fig.~\ref{fig:dimers_theta_fits}(a)
are plotted over the parameterized
curve obtained from these fits.
The same fit also is reproduced as
a dashed curve in Fig.~\ref{fig:dimers_theta_fits}(b),
where it is compared with experimental
data on comparable spheres.

\subsection{Effective-Sphere Measurement of Axial 
Position}

The same effective-sphere analysis
used to obtain a particle's effective
diameter and effective refractive index also
yields its axial position, $z_p$, relative
to the microscope's focal plane.
For micrometer-scale
colloidal spheres, the measured value of $z_p$ 
is found to have a precision of \SI{\pm 5}{\nm}
over the entire accessible axial range
\cite{krishnatreya_measuring_2014}.

\begin{figure}[h]
    \centering
    \includegraphics[width=0.8\columnwidth]{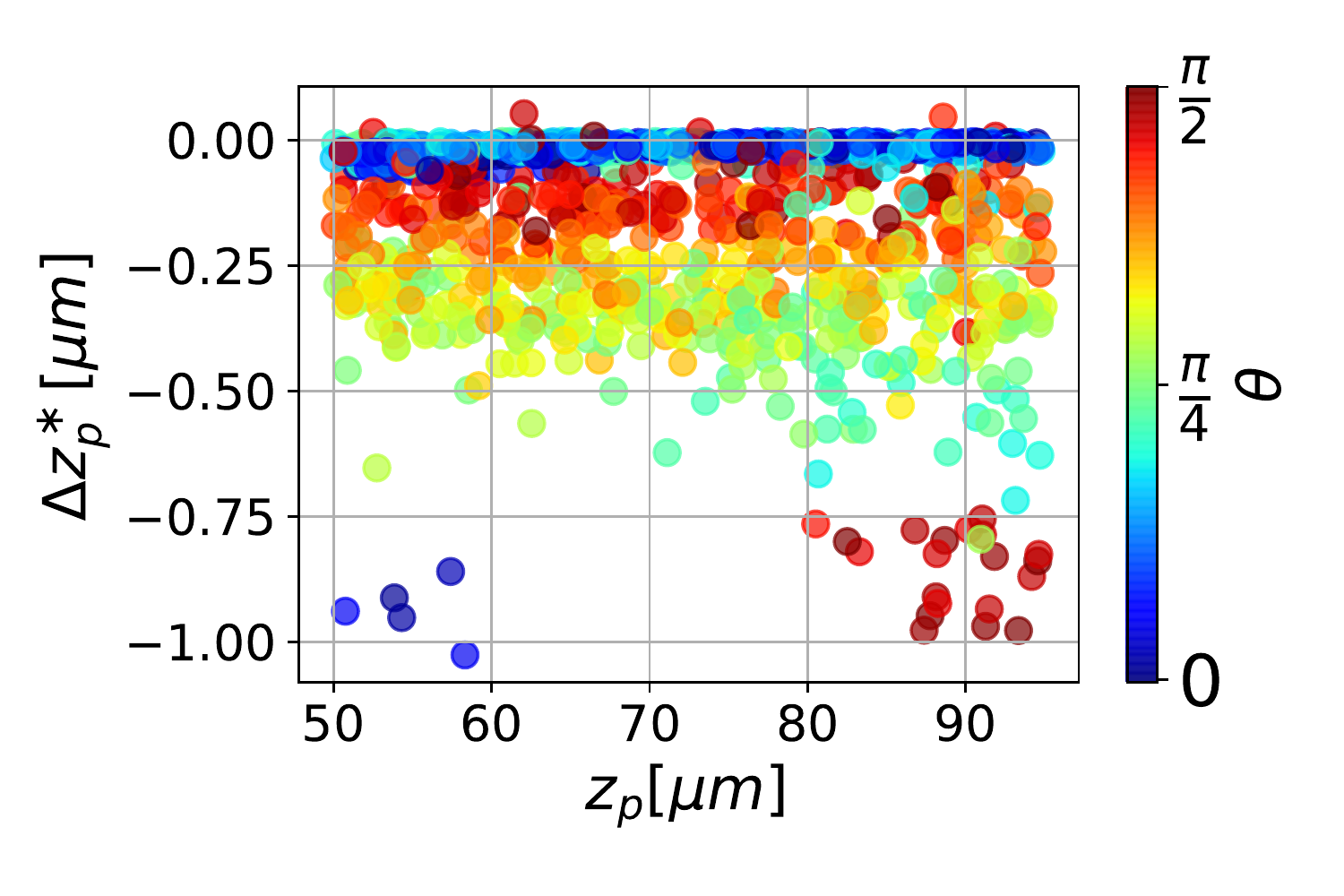}
    \caption{Error in the effective-sphere axial position,
    $z_p^\ast$, for the simulated symmetric dimers from Fig.~\ref{fig:dimers_theta_fits}(a) as a function of
    true axial position, $z_p$. Each point is colored by
    the dimer's orientation, $\theta$. Effective-sphere
    analysis tends to underestimate $z_p$, with the
    errors tending to be greater for dimers aligned at
    $\theta = \SI{45}{\degree}$ from the imaging plane.}
    \label{fig:zerror}
\end{figure}

Figure~\ref{fig:zerror} shows that effective-sphere
analysis yields surprisingly good values for
symmetric dimers' effective positions, with errors
consistently smaller than \SI{1}{\um} over a range
extending to nearly \SI{100}{\um}.
These measurements tend to underestimate the dimers'
height above the imaging plane
by an amount that increases with height
and depends on the dimers' orientations.
Errors tend to be smallest for dimers aligned
with the imaging plane or aligned with the optical
axis, and largest for $\theta = \SI{45}{\degree}$.
Outliers from these trends are likely to represent
convergence of the nonlinear least-squares fit
into a secondary minimum of the error surface
\cite{ruffner_2018}.

\section{Application to
Experimental Data} 

\subsection{Identifying Dimers}

When analyzing experimental holographic
characterization data such as the results
in Fig.~\ref{fig:schematic}(b), the identities of
individual particles are not known \emph{a priori}.
Sharp peaks in the $(d_p^\ast, n_p^\ast)$ plane
are likely to represent populations of monodisperse
spheres.
In this particular case, the sample is known to
include a population of \SI{1}{\um}-diameter
silica spheres coated with antibodies 
for a medical diagnostic test.
The band of low-index objects extending 
to the lower size limit of the instrument
are likely to represent 
protein aggregates \cite{wang_holographic_2016-1}
and 
bacterial contaminants
such as \emph{E. coli} \cite{cheong2015rapid}.
Based on the foregoing discussion of colloidal
dimers, we propose that the remaining broad
cluster of objects represents clusters
of the monomers, including symmetric dimers.

We automatically and objectively distinguish 
the three populations of particles by segmenting
the data into three categories using
the Gaussian mixture model, 
as implemented in \texttt{scikit-learn}
\cite{scikit-learn}.
Figure~\ref{fig:dimers_theta_fits}(b)
reproduces data from the monomer
and cluster categories,
with violet points corresponding to monomers and rainbow-colored 
points corresponding to clusters.
The third category, tentatively identified with
protein aggregates and bacteria, has been omitted.

After segmentation, we find that the
monomers have population-average properties
$d_p = \SI{1.054(9)}{\um}$ and $n_p = \num{1.435(5)}$.
These values were used for the simulations
that are plotted in Fig.~\ref{fig:dimers_theta_fits}(a).
The dashed curve in Fig.~\ref{fig:dimers_theta_fits}(b)
shows the distribution of
effective-sphere diameters,
$d_p^\ast(\theta)$, 
and refractive indexes, $n_p^\ast(\theta)$,
predicted for symmetric colloidal dimers.

As anticipated, the data points
identified with colloidal clusters are distributed
along the numerically predicted curve
for symmetric dimers.
This validates the assignment of those
data points as dimers.
This capability can be put to good use in
counting the numbers of monomers and
dimers in a sample, which in turn can be used
to measure the rate of dimerization.

The predicted dimer curve is
parameterized by polar angle, $\theta$.
We can assign a value for $\theta$ 
to each measured data point by identifying
the closest point along the predicted curve.
The data points for identified colloidal clusters
are colored accordingly in Fig.~\ref{fig:dimers_theta_fits}(b).
The measured orientations are not
uniformly distributed along the predicted
curve, but rather are concentrated
near $\theta = 0$.
This trend is expected for rod-like 
particles tumbling in a shear flow,
and is addressed in
Sec.~\ref{sec:jeffery}.
We first assess the nature of the
fluid flow in the microfluidic channel
and the particles' spatial distribution
within that flow.

\subsection{Axial Coordinates and Flow Profiling}

\begin{figure*}
    \centering
    \includegraphics[width=0.9\textwidth]{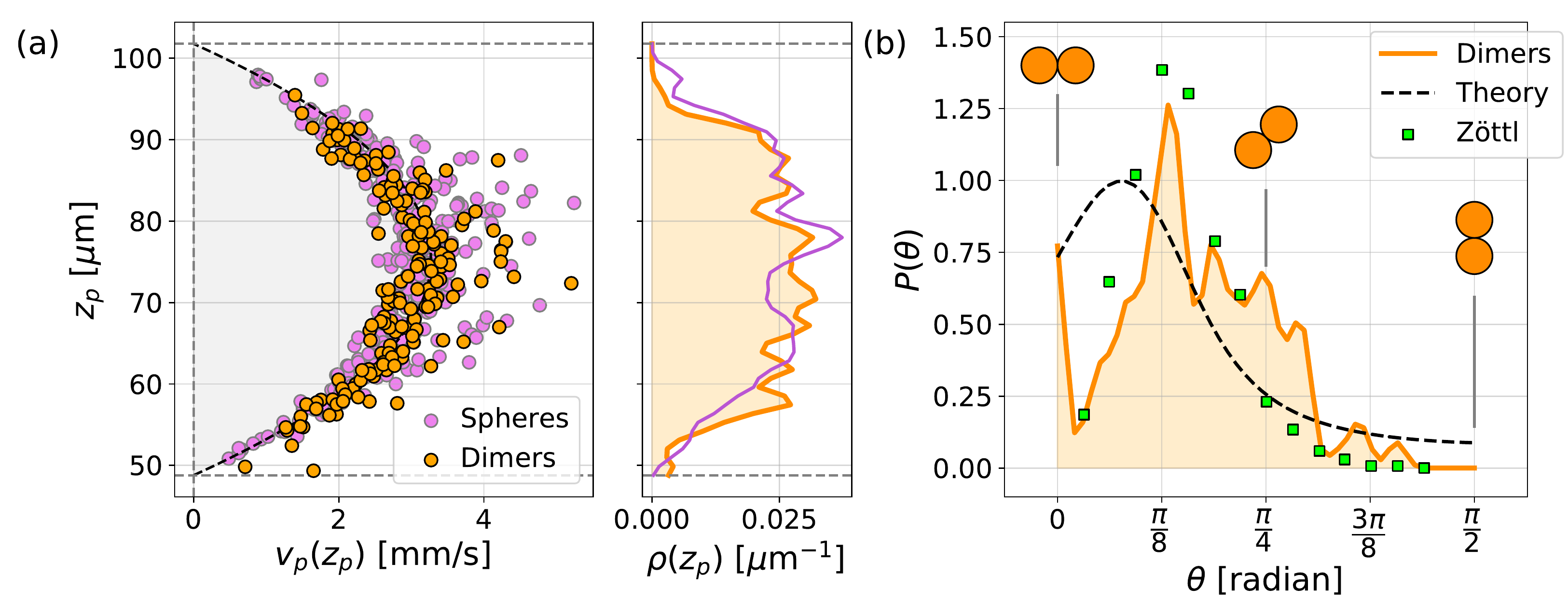}
    \caption{(a) The holographically measured flow profile,
    $v_p(z_p)$, for particles identified as dimers
    agrees with the flow profile for
    individual spheres co-dispersed in the same system.
    This agreement validates the
    effective-sphere estimate
    of $z_p$ for the dimers.
    The dashed curve is a fit to the
    parabolic profile expected for Poiseuille flow,
    which also yields estimates for the
    axial positions of the upper and lower walls, indicated by dashed lines.
    The projected distribution of axial positions, $\rho(z_p)$, shows that spheres and dimers are both concentrated near the midplane
    of the channel.
    (b) Angular orientation distribution, $P(\theta)$, for dimers (orange)
    estimated by projecting measured
    effective-sphere characterization data
    onto the computed curves for
    $d_p^\ast(\theta)$ and $n_p^\ast(\theta)$.
    Discrete (green) points reproduce experimental
    results on Brownian rods in a
    Poiseuille flow from \citet{zottl2019dynamics}.
    The dashed curve is the prediction
    of Eq.~\eqref{eq:prob} for
    $\Gamma_{max} = \num{50}$.}
    \label{fig:rho_theta}
\end{figure*}

Holographic characterization measurements yield
three-dimensional trajectories of colloidal particles
that can be used to map flow profiles in the
fluid medium \cite{cheong_flow_2009}.
The data in Fig.~\ref{fig:rho_theta}(a)
show the measured flow speed, $v_p(z_p)$, as
a function of the particles' axial positions, $z_p$,
for the experimental data from Fig.~\ref{fig:schematic}(b).
The data points are colored by their
classification as either monomers or dimers
from Fig.~\ref{fig:dimers_theta_fits}(b).
Being co-dispersed in the same fluid, both populations
of particles trace out the same parabolic profile
characteristic of Poiseuille flow in a rectangular
channel.
Fitting to a parabola yields estimates for the
position of the upper and lower walls of the channel,
which are plotted as horizontal dashed lines in
Fig.~\ref{fig:rho_theta}(a).
The $H = \SI{53}{\um}$ 
separation between the
walls is consistent with specified dimensions
of the xCell microfluidic channel (Spheryx, Inc.)
used for this measurement.
The width of the measured distribution of
flow speeds reflects variations in the actual
flow speed encountered in xSight measurements.

The flow profile reported by dimers is
consistent with the flow profile mapped by the
monomers. This observation further validates
the conclusion drawn from simulation that
effective-sphere characterization of symmetric
dimers yields accurate values for the dimers' 
axial positions.
The fluid's true flow profile serves as an
absolute reference for both populations of particles.
The sub-micrometer systematic downward bias in the dimers'
axial position anticipated from
Fig.~\ref{fig:zerror} is too small to resolve
in the measured distributions of axial positions,
$\rho(z_p)$, plotted in Fig.~\ref{fig:rho_theta}(a).

Monomers and dimers appear with roughly equal
probability at all heights in the channel
suggesting that they are present in this sample
at roughly equal concentrations.
The concentrations of both classes of particles
decrease sharply within \SI{10}{\um} of
either wall presumably because of shear-induced
migration mediated by collisions
\cite{frank2003particle}.
Having mapped the shear flow in the
channel and the locations
of particles within that flow, we can
interpret the observed distribution
of dimer orientations.

\subsection{Orientation Distribution and Jeffery Orbits}
\label{sec:jeffery}


Aspherical particles tend to rotate in shear flows,
their orientation vectors tracing out trajectories
called Jeffery orbits \cite{jeffery1922motion}.
The combination of shear forces, inertia, viscous
drag and diffusion cause aspherical particles, such
as dimers, to undergo complex gyrations.
Minimizing drag favors trajectories that align
dimers with the axis of vorticity
\cite{einarsson2015}, 
which lies in the imaging plane for our system.
We expect, therefore, to see more dimers aligned
with $\theta = 0$ than with $\theta = \pi/2$,
which is consistent with the measured
orientation distribution, $P(\theta)$,
plotted in Fig.~\ref{fig:dimers_theta_fits}(b).
This probability density is obtained with
a Gaussian kernel density estimator for the
distribution of dimer orientation angles
obtained by projecting the effective-sphere
estimates for $d_p^\ast$ and $n_p^\ast$ onto
the computed curves for symmetric dimers,
$d_p^\ast(\theta)$ and $n_p^\ast(\theta)$.

Given the measured flow speed on the midplane
of $v_0 = \SI{3}{\mm\per\sec}$, 
the channel height of $H = \SI{53}{\um}$
and the depletion of particles near
the wall, we would expect dimers to
experience a maximum
effective shear rate of
$\dot{\gamma} \leq 2 v_0/H \approx \SI{100}{\per\second}$.
The rotational diffusion coefficient
for colloidal dimers is similar to that
of a prolate spheroid of 2:1 aspect
ratio. For dimers of micrometer-diameter
spheres diffusing in water, we therefore
expect \cite{kim2013microhydrodynamics,hoffmann20093d},
$D_r \approx \SI{0.4}{\per\second}$.
These estimates suggest that our
system has a maximum rotational
P\'eclet number
$\Gamma_{\text{max}} = \dot{\gamma}/D_r \approx \num{200}$.


The discrete points plotted in Fig.~\ref{fig:rho_theta}(b) are experimental
results for $P(\theta)$ reported by \citet{zottl2019dynamics}
for micrometer-scale colloidal rods
being transported in a flow geometry
very similar to ours.
The orientation angles in this case
were estimated with conventional microscopy
by measuring the projected lengths of
the colloidal rods.
In this case, $\dot{\gamma} \leq \SI{16}{\per\second}$ and
$D_r = \SI{0.21}{\per\second}$,
yielding
$\Gamma_{\text{max}} = \num{76}$.
One data point at $P(0) = \num{2.5}$
falls outside the plot range of
Fig.~\ref{fig:rho_theta}(b).
Despite the factor of three difference
in rotational P\'eclet numbers between
this study and ours,
the reported values for $P(\theta)$
agree remarkably well.

While quantitative agreement between
the two experimental data sets may be
coincidental, common features in the
shape of $P(\theta)$ highlight 
general aspects
of particles' tumbling trajectories that
are not yet fully explained \cite{zottl2019dynamics}.
In both data sets, for example, the distribution of
particle orientations does not decrease
monotonically from $\theta = 0$,
but rather has a deep minimum
between $\theta = 0$ and a second
peak around $\theta = \pi/8$.
This minimum is notably absent from the
hydrodynamic simulations reported by \citet{zottl2019dynamics}.

The Fokker-Planck equation describing
the rotational advection and diffusion of
a Brownian particle in a uniform shear flow
has a steady-state solution
\cite{srinin2019nonlocal},
\begin{subequations}
\label{eq:couette}
\begin{equation}
Q(\theta, \Gamma) =  
\frac{
 f(\pi, \Gamma) \, 
 \int_0^\theta f(\theta', \Gamma) \, d\theta' 
 + 
 \int_\theta^{\pi} f(\theta', \Gamma) \, d\theta'}{
 z(\Gamma) \, f(\theta, \Gamma)},
\end{equation}
where $f(\theta, \Gamma) = \text{exp}(\Gamma u(\theta))$
and
$u(\theta) = \frac{\theta}{2} - \frac{\text{sin} 2\theta}{4}$.
The normalization constant, $z(\Gamma)$,
can be determined numerically and
is approximately given by
\begin{equation}
    z(\Gamma) \approx 
    \frac{(\frac{3 \pi}{2})^2+\Gamma}{1+\Gamma}
    e^{\pi \Gamma}.
\end{equation}
\end{subequations}
Equation~\eqref{eq:couette} is appropriate for
a rod-like particle in the
uniform shear of a Couette flow.
The Poiseuille flow mapped
in Fig.~\ref{fig:dimers_theta_fits}(a)
has a shear rate that varies linearly from
roughly $\dot{\gamma} = \SI{200}{\per\second}$
near the walls to zero on the midplane.
So long as $\dot{\gamma}$ does not vary
appreciably over the dimensions of the particle,
however, the orientation distribution function
may be approximated as an average of the Couette
result.
If, furthermore, the dimers are distributed
uniformly in the flow, the average over $z$
can be replaced by an average over $\Gamma$,
yielding
\begin{equation}
\label{eq:prob}
P(\theta) 
= \frac{1}{\Gamma_{\text{max}}} 
\int_0^{\Gamma_{\text{max}}} Q(\theta, \Gamma) \, d\Gamma
\end{equation}

The dashed curve in Fig.~\ref{fig:rho_theta}(b)
is the prediction of Eq.~\eqref{eq:prob}
for $\Gamma_{\text{max}} = \num{50}$,
which is smaller than the rotational P\'eclet
numbers in either our study or by \citet{zottl2019dynamics}.
Like the experimental results, this distribution
is peaked near $\theta = \pi/8$.
Unlike the experimental results, the theoretical
distribution does not feature a peak at $\theta = 0$,
nor is the predicted peak in $P(\theta)$
as sharp as the
experimentally observed peak.
Lower values of $\Gamma_{\text{max}}$ emphasize
diffusion and yield flatter distributions than are
seen experimentally.
Larger values sharpen the peak, but move it to smaller
angles.

It is possible that these discrepancies could
be resolved by accounting for inertial effects
\cite{einarsson2015}
or steric interactions between the particles
and the walls.
Regardless of their origin, the experimental
observation of peaks in $P(\theta)$
lend credence to the proposal that
holographic particle characterization provides
reasonable estimates for the axial orientation
angle, $\theta$, of symmetric colloidal dimers,
even when analyzed in the effective-sphere
approximation.
Effective-sphere analysis may be useful, therefore,
in resolving persistent discrepancies between
the theory of shear-induced tumbling and
experimental observations.

Holograms, such as the example in Fig.~\ref{fig:schematic}(a)
also encode information on a dimer's azimuthal orientation
angle, $\phi$. That information cannot be extracted
in the effective-sphere approximation, however,
because fits to Eq.~\eqref{eq:b} impose azimuthal
symmetry.
Simulations such as the example in Fig.~\ref{fig:dimers_theta_fits}(a) 
confirm that effective-sphere
characterization results are independent of $\phi$, which helps to explain why measurements
of $\theta$ appear to be reliable.

\section{Conclusions}

We have shown that in-line holographic microscopy
images of symmetric colloidal dimers can be interpreted
with a generative model for 
ideal spherical scatterers to obtain
estimates for the dimers' effective sizes,
$d_p^\ast$, effective refractive indexes, $n_p^\ast$,
and axial positions relative to the focal plane, $z_p$.
The effective-sphere diameter and refractive index
are parameterized by a dimer's inclination, $\theta$,
relative to the imaging plane in a way that can
be computed using cluster T-matrix theory.
The computed parameterization, in turn, can be used to
measure 
the orientation of dimers in experimental data.

Effective-sphere analysis offers substantial benefits
for analyzing colloidal dispersions including
the availability of commercial instrumentation that can rapidly perform
particle-resolved holographic characterization
measurements on large statistical samples.
When this technique is applied to nominally monodisperse
colloidal dispersions, effective-sphere analysis
usefully distinguishes dimers and other clusters
from spheres by predicting the range of effective-sphere
properties expected for dimers.
Differentiating monomers and dimers in this way can
be used to monitor the rate of
aggregation in colloidal dispersions, with immediate
application to label-free medical diagnostic testing based
on detection of colloidal agglutination.

Within the distribution of identified dimers,
effective-sphere analysis yields remarkably
accurate results for the dimers' three-dimensional
trajectories and also, apparently, for their three-dimensional orientations.
These observations suggest that holographic
particle characterization will be useful for 
studying the transport properties of aspherical
colloids, including both rods and clusters,
without requiring detailed fits to 
more descriptive theories of light scattering.
This approach may prove fruitful for addressing
outstanding questions about aspherical particles'
tumbling in shear flows.
It also should be useful for microrheology
with anisotropic probe particles
and for studying the interactions between aspherical
objects at the micrometer scale.
The ability to track three-dimensional trajectories and
orientations in near-real-time also should be useful
for studies of active colloidal particles \cite{ma2015inducing}
and bacteria \cite{cheong2015rapid},
many of which are aspherical.

Planned extensions to this work include
studies of more general types of colloidal clusters
including asymmetric colloidal dimers,
colloidal trimers and higher-order
clusters.
Effective-sphere analysis also should
be useful for initializing
fits to more complete theories for
scattering by clusters
\cite{fung_imaging_2012,fung_holographic_2013}.
All such present and prospective applications
build on the foundation of
commercial and open-source instrumentation
and software for holographic 
particle characterization
to provide a new window onto the properties
and dynamics of aspherical colloidal
particles and their dispersions.

The data and 
open-source software used for this study 
are available online
at
\url{https://github.com/laltman2/Dimer_HVM}.

\section*{Conflicts of Interest}

F.C.C.\ is a co-founder and Chief Technology Officer
of Spheryx, Inc., the company that manufactures the xSight instrument
for holographic particle characterization used in this study.
D.G.G.\ is a co-founder of Spheryx, Inc.

\section*{Acknowledgments}

This work was supported principally by 
the RAPID program of the National Science Foundation through Award No. DMR-2027013. Additional funding was provided by the
MRSEC program of the National Science Foundation
through Award Number DMR-1420073.
The Spheryx xSight used for this study was purchased
by the NYU MRSEC as shared instrumentation.

We are grateful to Andrew Hollingsworth for providing the
custom-synthesized colloidal silica spheres that were used in this study.
We also are grateful to Alexander Y. Grosberg and
Aleksander Donev for helpful conversations regarding
Jeffery orbits.

%

\end{document}